\documentclass[a4paper,11pt]{article}
\pdfoutput=1 

\usepackage{jheppub} 

\usepackage[T1]{fontenc} 
\usepackage{upgreek}
\usepackage{siunitx}

\title{\boldmath Commissioning and Operation of the New CMS Phase-1 Pixel Detector}

\author{Weinan Si on behalf of the CMS Phase-1 Pixel Upgrade team}

\affiliation{University of California, Riverside, \\ Riverside, CA, USA 92521}
\emailAdd{weinan.si@cern.ch}

\abstract{The Phase-1 upgrade of the CMS pixel detector is built out of four barrel layers (BPix) and three forward disks in each endcap (FPix). It comprises a total of 124M pixel channels in 1,856 modules, and it is designed to withstand instantaneous luminosities of up to $2 \times 10^{34}$\,cm$^{-2}$s$^{-1}$. Different parts of the detector were assembled over the last year and later brought to CERN for installation inside the CMS tracker. At various stages during the assembly tests have been performed to ensure that the readout and power electronics and the cooling system meet the design specifications. After tests of the individual components, system tests were performed before the installation inside CMS. In addition to reviewing these tests, we also present results from the final commissioning of the detector in-situ using the central CMS DAQ system. Finally we review results from the initial operation of the detector first with cosmic rays and then with pp collisions.
}

\begin{document} 
\maketitle
\flushbottom

\section{Motivation}

The pixel detector has been playing a vital role in the CMS experiment \cite{a} at the Large Hadron Collider (LHC) hosted at CERN. Seated closely around the beam pipe, it provides accurate vertex position measurement and robust reconstruction of charged particle trajectories. The original pixel detector, hereafter referred to as the Phase-0 pixel detector, consists of a three-layer barrel detector (BPix) and two endcaps at forward regions (FPix), each of the two having a two-disk structure. It was designed to withstand a luminosity of $1\times10^{34}$\,cm$^{-2}$s$^{-1}$ and a bunch spacing of 25ns. As the luminosity of the LHC keeps increasing with the great improvement of the machine and its successful operation, the number of interactions per bunch crossing (known as pileup) will increase to more than 50. Thus the Phase-0 pixel detector would not be suited for operation under such harsh conditions due to limited readout bandwidth. For example, the innermost barrel layer would be expected to have a data loss of up to 16\% \cite{b}, which is not tolerable. To maintain and even improve the tracking performance of CMS until the end of LHC Run 3, an upgrade of the pixel detector was performed, and the new Phase-1 pixel detector was installed in CMS in March 2017 during the extended year-end technical stop.

\section{CMS Phase-1 Pixel Detector}

This section focuses on the general upgrade and DAQ system.

\subsection{CMS Pixel Phase-1 Upgrade}

Compared with the Phase-0 pixel detector, the Phase-1 pixel detector features four major improvements.\\
First, with the detector volume mostly unchanged, one additional layer is added in the BPix and one additional disk is added for each endcap. This introduces 87.8\% more pixels increasing the number from 66 million to 124 million. Meanwhile, the radius of the innermost layer decreases from \SI{44}{mm} to 29\,mm, and the four-hit coverage within the tracking volume up to |$\eta$|=2.5 is guaranteed.
\\
Second, a new CO$_{2}$ cooling system was introduced in Phase-1 pixel detector. For a lower coolant mass flowing inside tiny stainless steel tubes, the novel two-phase CO$_{2}$ cooling system has been used to replace the old liquid C$_{6}$F$_{14}$ cooling system. Together with the ultra-light carbon fiber support frame and the movement of the auxiliary electronics from inside to outside the tracking volume, this greatly reduces the material.
\\
Third, the new Read-Out Chip (ROC) and the new Token Bit Manager (TBM) \cite{f}. To cope with the higher data rate, new ROCs have been developed and used. BPix layer 2-4 and FPix are using the  \texttt{psi46dig} chip \cite{c}, which is the upgrade version of the \texttt{psi46} chip \cite{d} that was used in the Phase-0 detector. The pixels are organized in 26 double columns, each with 80 rows. The hit address buffer is increased from 32 to 80 cells, the time stamp buffer is increased from 12 to 24 cells, and a global read-out buffer has been added. Moving from 40\,MHz analog readout to 160\,Mbit/s digital read-out, the \texttt{psi46dig} has an on-chip ADC to read out the signal height. Besides, the threshold for firing a pixel is reduced from 3500 electrons to as low as 1500 electrons. For the BPix layer1, the \texttt{PROC600} ROC \cite{e} which features a Dynamic Cluster Column Drain architecture is used to bear the higher hit rate of the innermost layer. To fit in with the different read-out design of the different layers or disks, various types of TBMs are used: FPix and BPix layer 3\&4 use \texttt{TBM08c}, BPix layer 2 uses \texttt{TBM09c}, and BPix layer 1 uses \texttt{TBM10}.
\\
Last, the new $\upmu$TCA based DAQ system. This will be discussed in more detail in the next section.

\subsection{$\upmu$TCA DAQ System}

\begin{figure}[tbp]
\centering 
\includegraphics[width=.8\textwidth,trim=0 0 0 0]{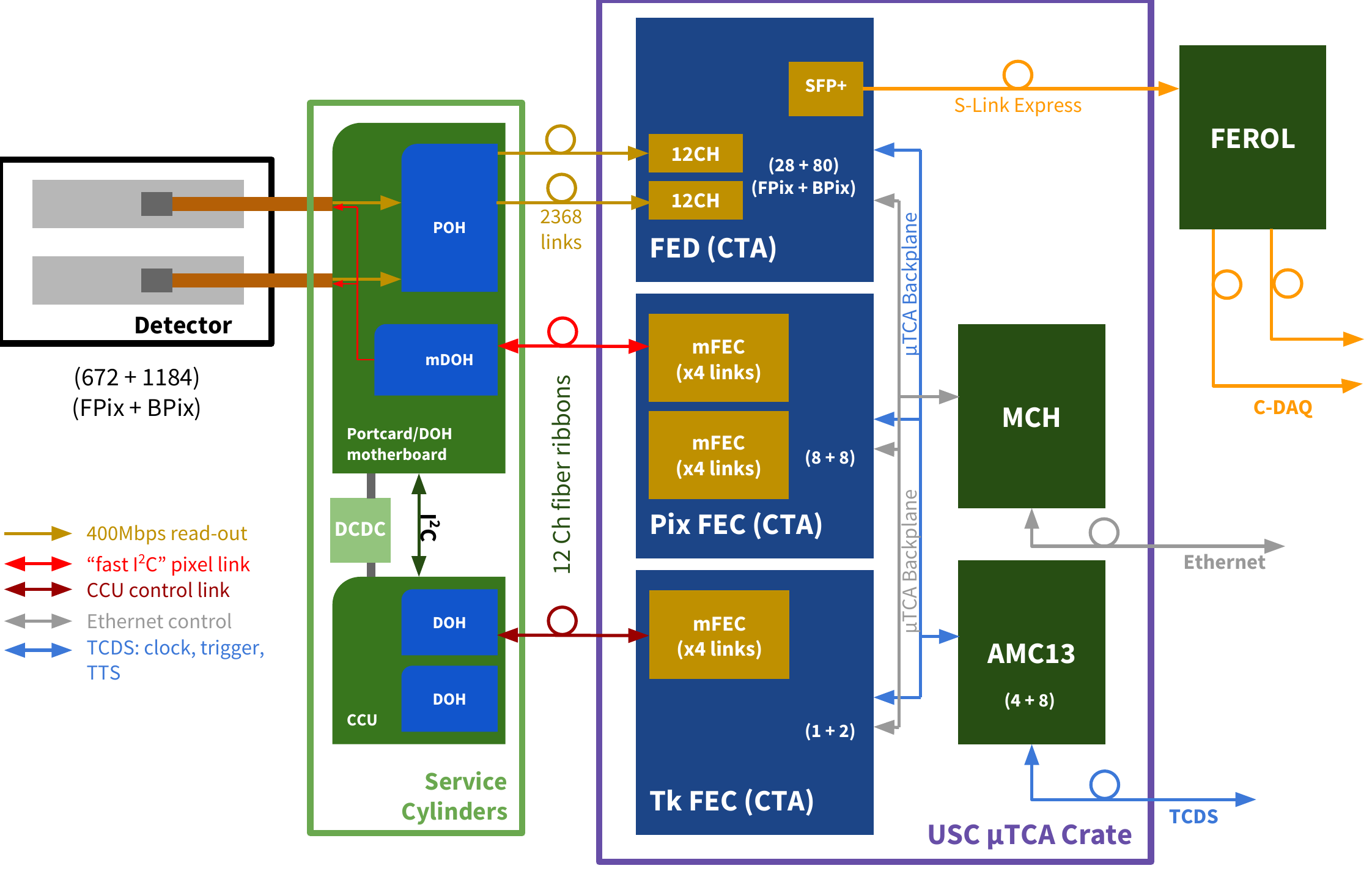}
\caption{\label{fig:daq} Complete overview of the $\upmu$TCA DAQ system of the Phase-1 pixel detector.}
\end{figure}

The core components of the pixel $\upmu$TCA DAQ system \cite{g}, as sketched in Fig.\ref{fig:daq}, are the three different flavors of boards which are built based on a common AMC card that features a Xilinx Kintex 7 FPGA and 4Gb of DDR3 RAM. Mounted with different types of mezzanine cards and loaded with different versions of firmware, they are turned into different flavors. In total, the pixel DAQ uses 3 Tracker Front-End Controllers (tkFEC), 16 Pixel Front-End Controllers (pxFEC) and 108 Front-End Drivers (FED). tkFEC programs auxiliary components of the pixel supply electronics such as opto-hybrids and DC-DC converters via the I$^{2}$C interface and PIA port of a Central Control Unit (CCU). The pxFEC distributes the clock, trigger and fast signals to the pixel modules, and programs the DAC registers of the ROC and the TBM. FEDs decode the incoming data stream from detector front-end, assemble data of 24 fibers into event fragments, and then push them to the central CMS DAQ system.

\subsection{Detector Read-out and Control}

The behaviour of the hardware is controlled by the Pixel Online Software (POS). Commands are sent through the ethernet from the PC to the MicroTCA Carrier Hub (MCH) and then distributed to the backend of the $\upmu$TCA boards.

\paragraph{Control} The tkFEC programs the auxiliary electronics including: a. DC-DC converters, which are introduced in Phase-1 and are used to power on pixel modules. b. The delay25 chip, which is used for data and clock alignment for the pxFEC. c. The TPLL (Tracker Phase-Lock Loop) and QPLL (Quartz Phase-Lock Loop), which are used for decoding the trigger and the clock coming from the TCDS (Trigger and Command Distribution System). The pxFEC programs the front-end ASICs including the TBM settings and the ROC settings.

\paragraph{Read-out} Pixel hits' information are firstly cached in the buffer waiting for the trigger and token acknowledgement. Once the trigger is received, a set of ROCs will be read out on a 160 MHz clock sequentially driven by token passing. Two data streams will go into the TBM, multiplexed in a 4-to-5 encoding scheme such that a 400 MHz data stream will be sent out through electrical cables to boards that carry the opto-hybrids. In these boards, the electrical signals are converted to optical signals and transmitted to the FEDs through the optical links. At last, FEDs collect data from 24 fibers (48 channels) and build the event fragments, then transmit them to the central CMS DAQ system.

\section{Detector Calibration Examples}

A lot of the software has been recycled and extended since much of the functionality of Phase-1 was already present in the Phase-0 detector. Compared with the Phase-0 pixel detector system, the differences are mainly on the read-out electronics. Therefore, two new calibrations are introduced in Phase-1 as follows:

\subsection{POH Bias Calibration}

\begin{figure}[tbp]
\centering 
\includegraphics[width=.65\textwidth,trim=0 0 0 0]{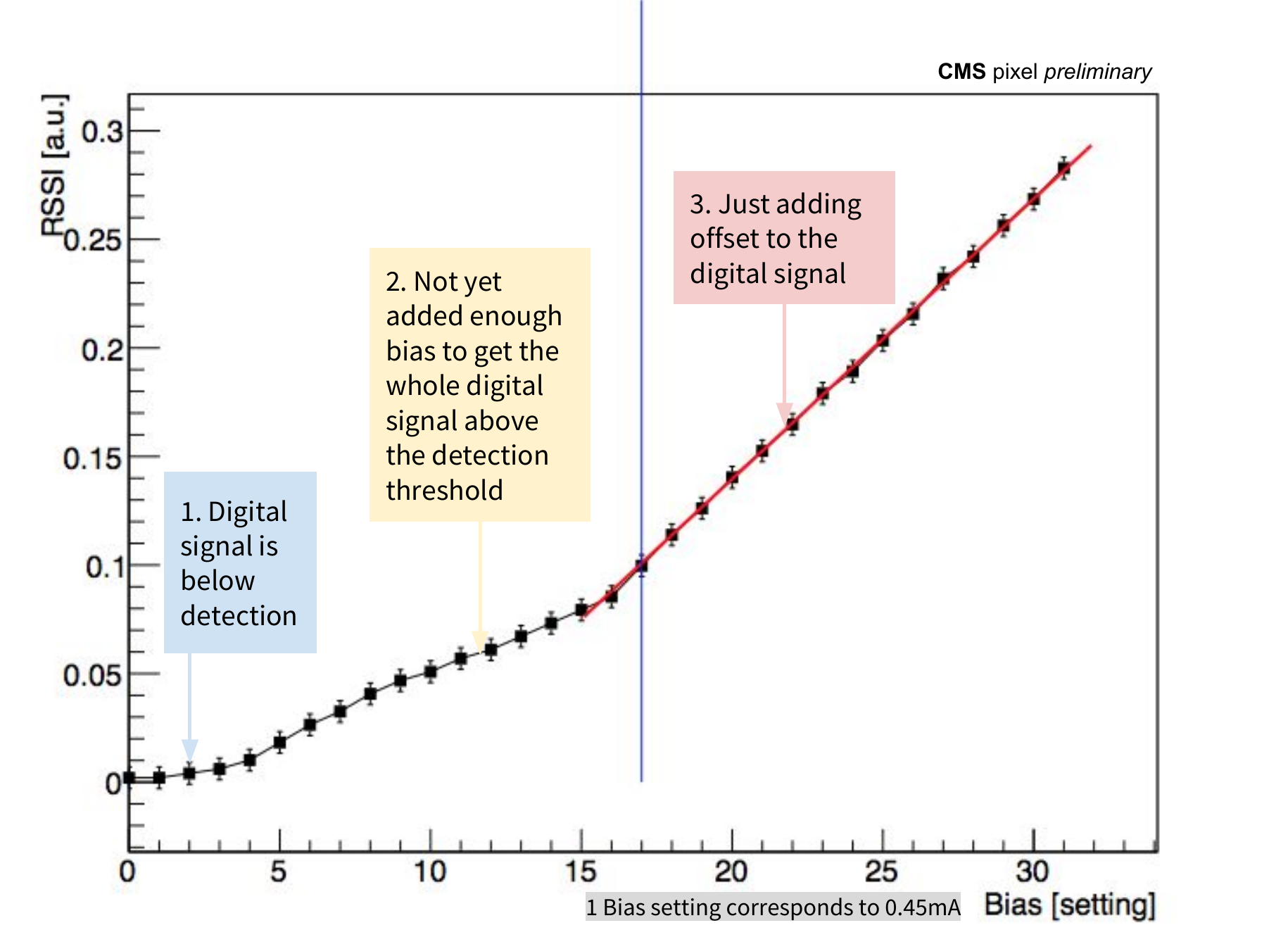}
\hfill
\caption{\label{fig:pohbias} POH bias calibration result example.}
\end{figure}

The POH (Pixel Opto-Hybrid) bias controls the amount of light sent from the detector to the pixel DAQ. As the POH bias increases, more light is sent, and the RSSI (Received Signal Strength Indication) value on the FED also increases. Three stages can be observed. In the first stage, the digital signal is below detection, the RSSI barely changes. In the second stage, the RSSI is growing slowly since there is not yet enough bias added to get the whole digital signal above the detection threshold. After a certain point, it goes into the third stage, where the offset is just being added to the digital signal. The bias value of the laser diode is chosen right after the second slope change, as indicated in Fig.\ref{fig:pohbias}. 

\subsection{TBM Phase Scan}

\begin{figure}[tbp]
\centering 
\includegraphics[width=.65\textwidth]{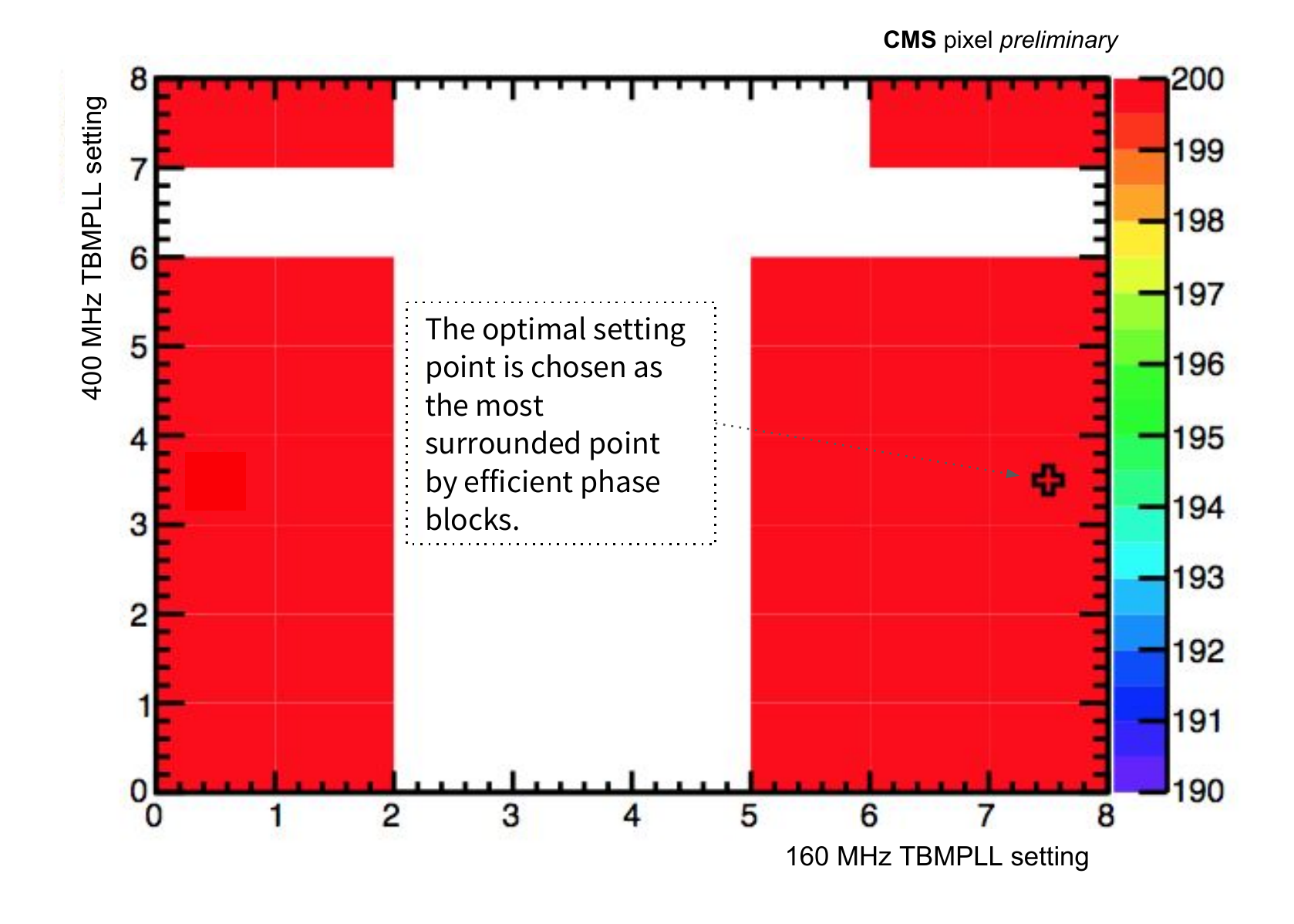}
\caption{\label{fig:tbmscan} TBM phase scan result example. The score is shown in the color scale, 200 is the "perfect" score, the white regions are phase settings who have underflow scores.}
\end{figure}

For each of the 48 channels of the FED, this calibration scans over all possible 400 MHz and 160 MHz TBM delay setting phases. For each of the setting combination points, a "score" is calculated based on the data stream structure. The optimal setting point is chosen as the point that is most surrounded by efficient phase blocks, as shown in Fig.\ref{fig:tbmscan}.

\section{Detector Commissioning and Operation}

The commissioning and operation of the Phase-1 pixel detector can be divided into three phases chronologically as follows.

\subsection{High (Random) Trigger Rate Test}

The purpose of this test is to test the pixel DAQ system. To test the capability of the FEDs, an emulator of optical pixel module output called the \texttt{FEDTester} has been developed. In the workbench setup, the \texttt{FEDTester} is connected with the CMS trigger system and provides input for the FEDs. By sending 100 kHz random triggers (the L1 rate in CMS), we loaded the FEDs with the emulated hits and read them out through the 10 Gbps link of the FEROL (Front-End Read-out Optical Link). The trigger rates will get throttled according to the FED status. In the production crate, we use the FED internal emulator to load 3 emulated hits per ROC in all FEDs, which is equivalent to a pileup of 105, and the FEDs can handle this situation with no problem.


\subsection{Cosmic Data-taking}

After the detector installation and the initial check-out, the pixel detector was turned on and took cosmic data. The initial cosmic data-taking was important to detect and solve obvious problems, and to bring the detector into a good state. For example, several timing setting scans were performed to get the optimal hit on-track efficiency. The noisy pixels were found and masked. Meanwhile, the "Private Resync" was devised to reduce the dead time caused by the pixel detector. In the Private Resync, the TCDS sends a resync command only to the pixel DAQ so that the other subsystems won't listen to it, and therefore considerable time can be saved. Besides, clean, low-occupancy muon tracks provide valuable input for the detector alignment and the hit residuals are improved by an order of 10. The optimal settings which were retrieved in the cosmic data-taking era provide a good starting point for the following collision data-taking.

\subsection{pp Collision Data-taking (Early Stage, till LHC TS1)}

\begin{figure}[tbp]
\centering 
\includegraphics[width=.5\textwidth,trim=0 0 0 0]{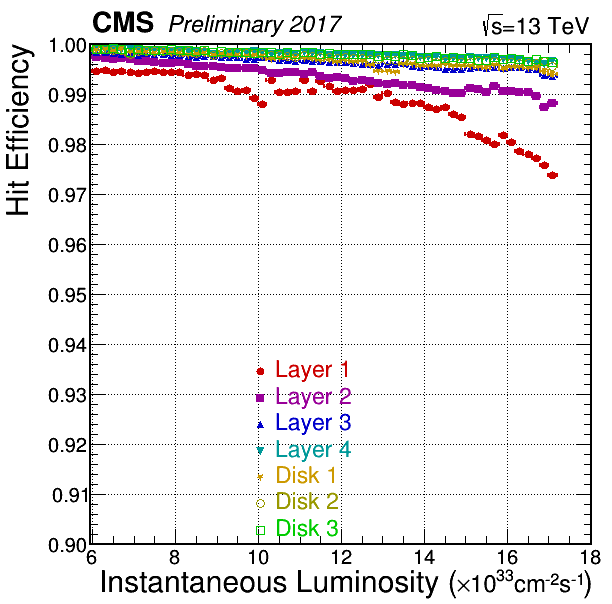}
\hfill
\caption{\label{fig:efficiency} Single hit efficiency \emph{vs} instantaneous luminosity. Data taken with the Phase-1 pixel detector during 2017.}
\end{figure}

On May 23rd, 2017, LHC delivered the first stable beam after the Phase-1 pixel detector installation. In the first timing scan, the data stream from layer 1 was observed to shift with respect to layer 2 by approximately half a clock cycle. An investigation was started on the timing shift between the \texttt{PROC600} and the \texttt{PSI46dig} since the BPix layer 1 modules are using the \texttt{PROC600} and the layer 2-4 and FPix modules are using the \texttt{PSI46dig}. The time alignment of layer 1 and layer 2, which shared a common programmable time delay, was difficult due to a faster layer 1 ROC. By the time of the LHC TS1 (Technical Stop 1), an optimal common plateau of efficiency with values close to 99\% was achieved for all the pixel layers and disks at a luminosity of $1.6\times10^{34}$\,cm$^{-2}$s$^{-1}$, as shown in Fig.\ref{fig:efficiency}. The setting was chosen to favor the layer 1 performance. Measurements are being taken to robustify the timing against possible issues as the LHC luminosity is increased and to even further improve the current layer 2 performance. Meanwhile, the FPix and BPix layer 3\&4 timing are well set and yield optimal performance.
\\
Another problem that was observed is a "stuck TBM", which means that the channel stops reacting to a trigger. This phenomenon depends on luminosity but not on the trigger rate. The most probable reason is Single Event Upset (SEU) on a new flip-flop in the new TBM. The affected channels could be recovered by power-cycling the module via the DC-DC converter. An automatic power cycle scheme was developed to cope with this situation.

\section{Detector Status and Conclusion}

Thanks to all the efforts in the building, component testing and calibration coming from many people, the Phase-1 pixel detector has been commissioned successfully and is taking data. The DAQ system is performing smoothly, and over 95.6\% of the detector channels are active and in good condition. Initial studies show that the performance of the more complex functions such as b-tagging, vertexing, and HLT electron reconstruction are already better than with the old pixel detector, which would not have been able to cope with the rates in the first place.

\end{document}